\documentclass{elsart}
\usepackage{graphicx}
\begin{document}

\begin{frontmatter}

\title{\bf Electronic Structure of LiMnO$_{2}$: A Comparative Study of the LSDA and LSDA+U methods }

\author{Nitya Nath Shukla}, 
\author{Rajendra Prasad\corauthref{cor}}
\corauth[cor]{corresponding author.}
\ead{rprasad@iitk.ac.in}
\address{Department of Physics, Indian Institute of Technology,
Kanpur-208016, India}

\begin{abstract}
A first-principles electronic structure study of orthorhombic, monoclinic and 
rhombohedral LiMnO$_{2}$ has been carried out
using the full-potential linearized augmented plane-wave 
method.
The exchange and correlations have been treated within the 
local spin-density approximation (LSDA) and the LSDA+U methods. In the 
LSDA, the stable ground state is antiferromagnetic insulator 
for the orthorhombic and monoclinic structures but is ferromagnetic metal for
the rhombohedral structure. The LSDA+U, on the other hand, predicts the ground state to be 
an antiferromagnetic insulator for all structures.
We find that strong correlations change the 
density of states dramatically around the Fermi level.
The LSDA+U predicts the nature of band gap to be a  mixture of charge transfer 
and Mn $d \leftrightarrow d$ like excitations for orthorhombic and monoclinic
LiMnO$_{2}$ and Mott-insulator for rhombohedral LiMnO$_{2}$ 
in agreement with the available experimental results.
The inclusion of U increases the magnetic moment on Mn and gives a value in
better agreement with experiment. However, Mn valency is not affected by 
the inclusion of U.
We have also calculated X-ray emission photoelectron spectra for the 
orthorhombic and monoclinic LiMnO$_{2}$ by the LSDA and the LSDA+U methods.
We find that LSDA+U gives better agreement with the available experimental results.

\end{abstract}

\end{frontmatter}

\section{ Introduction}
Transition metal oxides are interesting systems which exhibit phenomena such
as metal-insulator transition, magnetic phase transition and charge-ordering
etc \cite{imada,charge,charge1}. 
The electronic structure description of transition metal oxides (TMO), 
which are strongly correlated systems, is very important and has been a 
subject of many $\it{ab}$-$\it{initio}$ calculations on TMO
\cite{tera,bnd,vladi}. These studies 
are mostly based on the local spin density approximation (LSDA) 
\cite{parr,dreizler} and generalized gradient approximation (GGA) 
\cite{gga}. However, the LSDA and GGA fail in predicting many aspects 
of the electronic structure since the strong correlations between $d$ 
electrons play an important role. Thus inclusion of strong correlations
in the electronic structure calculations of TMO has been a challenging problem.

TMO are not only interesting from the basic physics point of view
but have a great potential for technological applications. In particular, 
the layered LiMnO$_{2}$ has attracted a lot of attention during the last decade
because of its potential use in rechargeable batteries 
\cite{tar,arm,pet}. 
LiMnO$_{2}$ in layered monoclinic structure was first synthesized by Armstrong 
${\it et} {\it al.}$ \cite{arm}, which triggered a lot of experimental
and theoretical activities on this system.
Detailed study on the structural stability of LiMnO$_{2}$ was carried 
out by Mishra and Ceder \cite{mishra}. They studied the effect of 
spin-polarization and magnetic ordering on the relative stability of various 
structures using the density functional theory (DFT) in the LSDA \cite{parr,dreizler} and GGA \cite{gga}. 
The LSDA calculation gives metallic ferromagnetic(FM) ground state for 
the rhombohedral LiMnO$_{2}$ but the GGA
calculation predicts antiferromagnetic(AFM) ground state.
Terakura ${\it et} {\it al.}$ \cite{tera} showed 
that the LSDA predicts metallic ground state for many transition metal oxides
whereas experimentally these oxides are found to be insulators. 
They also showed that when AFM ordering is 
included in the LSDA calculations, the band gap can open up although it 
turns out to be smaller than the experimental gap \cite{bnd}.
However, the LSDA calculation performed by Galakhov ${\it et} {\it al.}$ 
\cite{gala} on orthorhombic LiMnO$_{2}$ 
shows metallic AFM ground state. Also the
magnetic moment on Mn was found to be smaller than the experimental work \cite{ortho2}. 

LiMnO$_{2}$ can exist in various phases such as orthorhombic, monoclinic, 
spinel etc., the ground state structure being orthorhombic (Pmmn)
\cite{ortho1,ortho2}.
The favourable structure for rechargeable batteries is the layered rhombohedral 
(R$\bar{3}$m). 
But Mn$^{3+}$ is a Jahn-Teller active ion in
a rhombohedral environment and gives rise to a cooperative monoclinic 
distortion and as a result the rhombohedral phase is not stable. 
Doping of some elements such as Co can stabilize the rhombohedral 
structure which has been studied by Prasad ${\it el} {\it al.}$ 
\cite{prasad,prasad1} using the GGA. But they found the stability of structure at 
about 32 \% \cite{prasad2} doping of Co which is very high as compared to 
the experimental value of 10 \% \cite{arm1}. This can be attributed 
partially to the approximate treatment of correlation effects in the GGA. 

As discussed above, several studies 
\cite{mishra,gala,prasad,prasad1,prasad2}
indicate that neither the LSDA nor the GGA describes the electronic structure of 
LiMnO$_{2}$ satisfactorily. Thus for a better description of the electronic
structure of this material the electron
correlations between d-electrons must be included.
The LSDA+U method \cite{aniso1,ldau} provides a simple 
way of incorporating correlations beyond the LSDA 
and GGA by introducing orbital dependent potential. The advantage of the 
LSDA+U method is the ability to treat simultaneously delocalized conduction
band electrons and localized 'd' or 'f' electrons within the same computational
scheme. The LSDA+U improves the 
electron-energy-loss spectra and parameters characterizing the structural 
stability of NiO \cite{duda,aniso1} compared to the LSDA. Recent studies
by the LSDA+U on some other oxides show the 
improvement over the LSDA results \cite{vladi,ceder,roll,zhou}.

The aim of this paper is to study the effect of electron correlations on
the ground state electronic structure properties and magnetic properties of 
LiMnO$_{2}$ using the LSDA+U method \cite{aniso1,ldau}.
We hope that this calculation will be able to give correct ground state
structure for various phases of LiMnO$_2$ and will provide starting point for
future calculation with dopants.
The electronic structure calculations have been done
using the full-potential linearized 
augmented plane-wave (FLAPW) method \cite{wien}. 
We consider only orthorhombic, monoclinic and rhombohedral LiMnO${2}$.
Since the value of U is not known for LiMnO$_{2}$ and there is no
unique way of finding the value of U, we adopt the following procedure. 
We start our calculation for the 
orthorhombic structure with the value of U for MnO as given in the literature 
\cite{aniso}. 
We then repeat the calculations using different 
values of U for orthorhombic, monoclinic, rhombohedral structures.
The trends are similar with different values of U for orthorhombic, monoclinic
and rhombohedral structures. Our LSDA+U results show that the  
ground state of rhombohedral LiMnO$_{2}$ is an AFM insulator although the LSDA 
shows it to be a FM metal. 
The band gap and magnetic moment of Mn 
increase in the LSDA+U as compared to the LSDA. 
The calculation of band structure and density of states (DOS) shows that the strongly
correlated effects dominate over the crystal field effects. We have also
calculated X-ray emission spectra (XES)\cite{xes1,xes} using the LSDA 
as well as the LSDA+U for the orthorhombic and monoclinic structures. 
The LSDA+U results at U=8.0 eV are closer to the experimental results.

The organization of paper is as follows. In Sec. \textbf{II} we provide 
computational details of the present work. In Sec. \textbf{III} we present our 
results and discuss them. Finally, we give our conclusions in Sec. \textbf{IV}.

\section{Computational details}
\subsection{Method}

The LSDA and LSDA+U methods have been discussed in great details in Ref.
\cite{parr,dreizler,aniso1,ldau} and will not be discussed again.
Here we give only the relevant computational details.
We have used the full-potential linearized augmented 
plane-wave (FLAPW) method \cite{wien} in a 
scalar relativistic version without the spin-orbit coupling.
The exchange and correlation potential has been included within the LSDA of 
the density functional theory \cite{parr,dreizler}.
To treat strong correlations, the localization of the $d$-states has been
corrected by means of the LSDA+U method \cite{aniso1,ldau} 
which is the LSDA with an on-site Coulomb potential for the $d$-states. 
The double-counting term, which is already
included in the LSDA, has been subtracted as given in Eq. (4) of 
Ref.\cite{ldau}.
For the LSDA+U calculations as described in Ref. \cite{aniso1,ldau}, 
we need to know the Hubbard parameter U and the
screened exchange energy J which is almost a constant about 1 eV 
\cite{jvalue}. We started our calculations for all LiMnO$_{2}$ structures
with the value of U=6.9 eV and J=0.82 eV for MnO as given in literature 
\cite{aniso} and then repeated the calculations at different values of U 
without changing J. 
The XES spectra were calculated using the formalism of Neckel 
$\it{et}$ $\it{al.}$ \cite{xes1,xes}. According to this method, 
K- and L-spectrum intensities of an atom A have 
the following proportionality 
\begin{equation}
I_{K}(E) \propto {1 \over 3} E^3 M_{A}^2(p,1s,\epsilon) {\rho}_{p}^A(\epsilon) 
\end{equation}
\begin{equation}
I_{L}(E) \propto E^3 (M_{A}^2(s,2p,\epsilon) {\rho}_{p}^A(\epsilon) + {2\over5} M_{A}^2(d,2p,\epsilon) {\rho}_{d}^A(\epsilon))
\end{equation}
where $E=\epsilon -\epsilon_{core}$ is the energy of the
emitted x-ray with $\epsilon$ and $\epsilon_{core}$ being the energies for 
valence and core electrons respectively. $\rho_{l}^A(\epsilon)$ is the 
$l$-component of the local (inside the atomic sphere A) 
DOS and $M_{A}^2$ is the radial transition probability. The 
calculated spectra include broadening for the spectrometer and lifetime 
broadening for core and valence states. For Mn spectrum, we have used 0.9 eV 
and 0.75 eV broadening for core and valence lifetime and 0.5 eV for 
spectrometer while these parameter for O spectrum are 0.4 eV, 0.65 eV and 0.2
eV. The integration 
over the Brillouin zone in the 
self-consistency cycle is performed using 64, 100, and 113 k-points in 
irreducible Brillouin zone for the orthorhombic, rhombohedral and monoclinic 
LiMnO$_{2}$ structures respectively. An improved tetrahedron method has been
used for the Brillouin-zone integration \cite{blochl}.
 The maximum $l$ value in the radial sphere expansion is $l_{max} = 10$, and the
largest $l$ value for the non-spherical part of the Hamiltonian matrix is
$l_{max,ns} = 4$. The cutoff parameters are $R_{mt}K_{max} = 7$ for the plane
wave and $R_{mt}G_{max} = 14$ for the charge density. 
The number of plane waves ranges from 11510 to 20853, depending on 
the structure of LiMnO$_{2}$.

\subsection{Structure}

The calculations are performed for orthorhombic, monoclinic and rhombohedral 
structures with space group Pmmn, C2/m and R$\bar{3}$m respectively. The
experimental lattice parameters are available for orthorhombic and monoclinic
structures. Therfore, we have used 
experimental lattice constants for the orthorhombic \cite{ortho1} and
monoclinic \cite{arm} structures . The rhombohedral structure is not a stable
structure  and its experimental lattice constants are not available.
Therfore,
the optimized lattice constants are used for the rhombohedral structure 
as given by Mishra and Ceder \cite{mishra}. 
For the AFM and FM calculations,
we construct a supercell of 16 atoms for the orthorhombic LiMnO$_{2}$.
The AFM ordering is given in Fig. 1 for the orthorhombic structure. 
To get the supercell, we double the unit-cell of 8 atoms in the $\bf{x}$-
direction and arrange Mn atoms in AFM ordering along the 
$\bf{x}$-direction and in FM ordering along the $\bf{y}$-direction in both planes (z=0.4
and z=0.6). We have also checked our calculation using a bigger supercell of
32 atoms but the band gap and magnetic moments do not change significantly. 
On the other hand, the
computational time increases drastically with increasing number of atoms in 
supercell. Thus we have chosen a supercell of 16 atoms for our calculations 
while 8 atoms cell have been used in Ref.\cite{gala}. 
For the monoclinic and rhombohedral structures 
the AF3 ordering has been used for the AFM case as suggested by Singh 
\cite{singh} .
We have followed the construction of AFM structures 
as given by Prasad ${\it et} {\it al.}$ \cite{prasad1} for 
the monoclinic and rhombohedral structures. 

\section{Results and Discussion}

We have calculated the total energy for spin-polarized and non-spin-polarized 
configurations of the orthorhombic, monoclinic and rhombohedral LiMnO$_{2}$ 
structures using the LSDA and LSDA+U methods. 
In all cases we find that the non-spin-polarized configuration has higher energy
 than the spin-polarized configurations. Therefore,
in the following, we shall discuss our results
for only spin-polarized configurations for various structures.
We will discuss the effect of different U values on 
electronic structure, magnetic moments and band gaps of these structures.
\subsection{Electronic Structure}
\subsubsection{Orthorhombic LiMnO$_{2}$}

The calculations have been done using the FM and 
AFM ordering of Mn in the orthorhombic structure. 
The AFM configuration has lower energy in both the LSDA 
and the LSDA+U calculations. As the value of U is not known, we begin with 
U=6.9 eV, which is the value for MnO as given in the literature 
\cite{aniso} and then repeat the calculations for different values of U.

Fig. 2 shows the calculated total and partial DOS for Mn and oxygen in the AFM 
configuration by the LSDA and the LSDA+U methods (for U=8.0 eV). The O-2p DOS is shown for
the oxygen at two different planes of 2b site (O1 at z=0.13 and O2 at z=0.60).
We shall first discuss the LSDA DOS which is shown in the panel (a). We see that
the total DOS (top curve in left) in the occupied region near E$_{F}$ 
(0.0 to -3.0 eV) has a large contribution from Mn-d states as is evident 
from the partial DOS of Mn-d and O-2p states. 
The O-2p DOS mainly contributes in energy range -7.0 eV to -3.0 eV.
The partial DOS of O-2p states at two different sites shows slight difference 
because the Mn-O bond lengths are different for these sites. 
We note that the crystal field splits Mn d-t$_{2g}$ and d-e$_{g}$ DOS giving
rise to a band gap of 0.30 eV. 
This is in contrast to the result obtained by 
Galakhov ${\it et} {\it al.}$ \cite{gala} using LMTO-ASA method 
\cite{lmto} which shows the ground state to be 
a metallic AFM. This
difference can be attributed to the difference in magnetic ordering, 
lattice parameter and atomic
positions as well as the FLAPW method which treats potentials more accurately 
than the
ASA. We have taken the AFM ordering along $\bf{a}$ direction while it is along
$\bf{b}$ direction in Ref. \cite{gala}. The magnetic ordering also
affects the magnetic moment of Mn \cite{singh} which is higher 
in the work of Galakhov ${\it et} {\it al.}$ (3.45 $\mu_{B}$)
\cite{gala} as compared to our results (3.29 $\mu_{B}$). 
In the AFM ordering, the magnetic
moment on Mn is 3.29 $\mu_{B}$ and each O sphere also gets polarized with a
 moment 0.05 $\mu_{B}$, while in FM ordering, the Mn has moment 3.43 $\mu_{B}$.
The magnetic moment of Mn in the AFM state is smaller than the experimental 
value \cite{ortho2} of 3.69 $\mu_{B}$.

Now we discuss the LSDA+U results which are shown in panel (b) of Fig. 2.
We see that the DOS in the occupied  
region from -3.0 eV to 0.0 eV has a large contribution from O-2p states and a 
small contribution from Mn-d states which is in sharp contrast to the LSDA
result. This is due to the
added electron-electron repulsion U, which pushes the minority bands 
up in energy by roughly U/2, and pushes the majority Mn-d bands down 
by roughly U/2. Thus 
the peaks in Mn d-t$_{2g}$ and d-e$_{g}$ DOS (majority bands) are shifted 
below E$_{F}$  by $\sim$4.0 eV.
Because of these shifts the O-2p DOS increases near E$_{F}$ as shown in the 
lower panels of Fig.2 (b) and is 
comparable to Mn d-e$_{g}$ DOS (second panel of Fig. 2(b)) near E$_{F}$. 
Thus the band gap is created between O-2p band and upper Mn-d band as
well as between Mn d-d bands. 
The opening of the gap between O-2p and Mn-d bands is a signature of a 
charge-transfer insulator \cite{charge}.
The shift in occupied majority spin Mn-d bands changes 
the band gap from Mn-d band splitting (crystal field effect) to a 
charge-transfer type O-2p and Mn-d gap with some contribution of Mn d-d like
excitations. Thus the LSDA+U 
predicts it to be a mixture of charge-transfer and Mn d$\leftrightarrow$d 
excitations like insulator in contrast to the LSDA, which
predicts it to be a band insulator. Thus the inclusion of U 
qualitatively changes the electronic spectrum. The LSDA+U prediction is in
agreement with experimental result \cite{gala} which shows
the top of valence band in LiMnO$_{2}$ to be dominated by O-2p states.
We also note that the band gap (1.95 eV) in the LSDA+U calculation is much
larger than that in the LSDA calculation (0.30 eV).
Thus we find that the strong correlation effects 
dominate over the crystal field effects. 

The magnetic moments of Mn for AFM case is given in Table I.
The U=0.0 eV value corresponds to the LSDA results. It is clear from 
Table I that Mn magnetic moment increases with increasing 
value of U and the experimental moment of Mn in AFM 
configuration is obtained at U=8.0 eV which is 3.69 $\mu_{B}$ compared to 3.29
$\mu_{B}$ without U. This shows that inclusion of U is important in these
systems. Mn magnetic moment also increases in FM configuration 
from the LSDA value of 3.43 $\mu_{B}$ to a value 3.71 $\mu_{B}$ at U=8.0 eV.
This is because in the LSDA+U, the electron occupation decreases in minority 
Mn-d states  and increases in majority Mn-d states (see Table II).

We have also calculated the difference in total energy per unit cell between 
the AFM and FM ordering  as a function of U. The difference between the 
energies of FM and AFM ordering decreases as we increase the value of U.
We find that the magnetic ground state is not affected by the choice of U 
value and remains antiferromagnetic for all values of U (see Table III).
 At U=8.0 eV, the AFM configuration has an energy
0.05 eV per formula unit-cell below the FM configuration 
while this energy difference is 0.30 eV in the LSDA (U=0.0 eV).
We have also calculated the band gap as a function of U 
and found that it too increases with increasing value of U.

\subsubsection{Monoclinic LiMnO$_{2}$}

Next we discuss the results for the monoclinic structure in FM and AF3
configurations.
We have carried out calculations at
different values of U similar to the orthorhombic case.
The LSDA as well as LSDA+U, predict insulating state for both the magnetic 
orderings in monoclinic structure.
The AF3 configuration has lower energy compared to the FM configuration in 
both the LSDA and the LSDA+U calculations. 
In Fig. 3, we show the calculated DOS for the AF3 ordering. Panel (a) shows the 
LSDA total DOS and the partial DOS for Mn d-t$_{2g}$, d-e$_{g}$ and O-2p states.
As in the orthorhombic case (and also rhombohedral case) Mn-d band splits in 
d-t$_{2g}$ and d-e$_{g}$ bands but there is further splitting of t$_{2g}$ and e$_{g}$ 
bands due to the cooperative Jahn-Teller distortion. 
The band gap arises due to this band splitting which is 0.61 eV.
Note that without the Jahn-Teller splitting the system would be a metal as 
in the rhombohedral case (see next section).
Mn-d bands contribute to the top of valence band (in energy range 0.0 eV
to -3.0 eV) with a small contribution
from O-2p bands, similar to the LSDA results of the orthorhombic case. 
The magnetic moment 
(see Table I) on Mn sphere is 3.29 $\mu_{B}$ in the AF3 configuration and 
each O also get polarized with a moment 0.05 $\mu_{B}$. In the FM ordering, the 
band gap is 0.17 eV and the magnetic moment on Mn and O are 3.42 $\mu_{B}$ 
and 0.13 $\mu_{B}$ respectively. We find good agreement between our LSDA 
results and the results reported by Singh \cite{singh}. 

We show the LSDA+U DOS (for U=8.0 eV) in panel (b) of Fig. 3.
The occupied Mn-d bands are shifted towards
lower energy (-4.0 eV to -7.0 eV) while the unoccupied Mn-d bands are shifted up
in energy because of added electron-electron repulsion U. 
This increases the band gap by 1.19 eV.  
We see that the O-2p DOS increases around E$_{F}$ in energy 
range 0.0 eV to -3.0 eV.
Hence, the top of O-2p bands and the Mn-d bands merge, and so there is no high 
intensity peak structure in the total DOS in energy range
0.0 eV to -3.0 eV as seen in the LSDA.
We note that the Mn d-t$_{2g}$ and O-2p DOS have sharp peaks close to E$_{F}$
and hence Mn-d states and O-2p states contribute significantly to the 
top of the valence band.  
Thus similar to the orthorhombic case, monoclinic LiMnO$_{2}$ is also
a mixture of charge-transfer type and Mn d$\leftrightarrow$d excitations
like insulator. The band gap changes to 1.80 eV
in presence of U while it is 0.61 eV in the LSDA.
Thus we find that the electron-correlation effects 
tend to wash out distinct crystal field effects similar to the  
orthorhombic case. 

Table I shows the magnetic moments and band gaps for different
values of U for AF3 ordering.
Mn  magnetic moment increases to 3.79 $\mu_{B}$ at U=8.0 eV.
Mn magnetic moment increases with U in the case of FM ordering also 
which is 3.67 $\mu_{B}$ at U=8.0 eV.
This increase in Mn magnetic moment is similar to orthorhombic case.
The AF3 configuration has lower energy by an amount 0.61 eV per formula unit compared to  
the FM configuration in LSDA and 0.17 eV per formula unit in LSDA+U 
(see Table III). 

\subsubsection{Rhombohedral LiMnO$_{2}$}

In this section, we present results for the rhombohedral case in FM and AF3
configurations.
Although this structure is unstable, it can be stabilized by doping with
Co, Mg etc. \cite{prasad,prasad1}. Thus it would be interesting to see
the effect of strong correlations on the electronic structure of rhombohedral
LiMnO$_{2}$ and compare with the earlier work \cite{mishra}.
Similar to the orthorhombic case, we have 
used different values of U and fixed value of J for the LSDA+U calculations. 
The magnetic moment of Mn and band gap increase with increasing
value of U for AF3 ordering.
The LSDA predicts FM configuration as a 
lower energy state while LSDA+U gives AF3 insulator as the lower energy state. 
Fig. 4 shows the calculated total DOS for AF3 and FM configurations by
the LSDA (panel (a) and LSDA+U (for U=8.0 eV) (panel(b)).
The LSDA predicts AF3 configuration to be a metal and FM configuration to be
a half metal. This can be seen from Fig. 4 which
shows finite DOS at E$_{F}$ for AF3 as well as FM configurations
(for minority spin) while the majority spin DOS shows gap at E$_{F}$.

In the LSDA+U calculation, the band gap opens up with a value 0.85 eV and 
3.18 eV at U=8.0 eV in AF3 and FM configurations respectively. 
The inclusion of U shifts the bands and opens up a gap at E$_{F}$. 
In the AF3 configuration, the top of valance band has major contribution from 
Mn-d states in energy range -1.0 eV to 0.0 eV while O-2p contribution is in
energy range -8.0 eV to -2.0 eV. The AF3 LSDA+U DOS are similar to monoclinic 
and orthorhombic LSDA DOS as it has strong Mn-d contribution at top of
valence band.
As the band gap is between Mn d-d bands, the insulating phase is a 
Mott-insulator. The nature of band gap is different between rhombohedral
and monoclinic, orthorhombic structures because of their geometry differences.
The Jahn-Teller distortion is responsible for the band gap in LSDA calculations of 
monoclinic and orthorhombic case while the inclusion of U has shifted the bands
and created band gap in rhombohedral case. 

In the LSDA, magnetic moment of Mn is 2.29 $\mu_{B}$ and 1.82 $\mu_{B}$ 
in AF3 and FM configuration respectively.
Also there is a small magnetic moment on oxygen. The magnetic 
moments and band gap are given in Table I for AF3 configuration.
Our LSDA results are in good agreement with earlier calculation 
\cite{mishra}. 
Magnetic moment of Mn has a large change compared to the 
monoclinic structure in AF3 ordering (see Table I) because the LSDA gives an 
insulating AF3 ground state in monoclinic structure but it gives metallic FM 
ground state in rhombohedral structure. 

The AF3 configuration has an energy 0.581 eV per formula unit below that of FM  
configuration in the LSDA+U and the FM configuration has an energy 
0.52 eV per formula unit below that of AF3 configuration in the LSDA (see
Table III). The 
energy difference in the LSDA+U is higher in this case as compared to the 
monoclinic and orthorhombic. This can be attributed to the difference in ground
states by the LSDA in these structures. 

Rather than presenting the full band structure we present 
a simple energy-level diagram which will help to understand our DOS results.
Fig.5(a) schematically shows the energy-levels for Mn 
surrounded by O in an octahedral environment when U is assumed to be zero. Mn-d
levels, in the crystal field shown in the left hand side (in Fig. 5(a)),
form bonding and antibonding levels after hybridization to O levels, as shown 
in the right hand side. The $dp\sigma$ levels arise from $\sigma$-bonding 
between Mn $e_{g}$ and O-2p orbitals while the $dp\pi$ levels arise from 
$\pi$-bonding between Mn $t_{2g}$ and O-2p orbitals. The $\sigma$-bonding is 
usually much stronger than the $\pi$-bonding \cite{mo}. 
Thus the $dp\pi$ level has smaller O-2p contribution.
Hence, the Mn d-d gap ($\Delta_{d-d}$) is smaller than O-2p 
Mn-d gap ($\Delta_{p-d}$) as shown in Fig. 5(a). 
This picture changes after the inclusion of U as shown in Fig. 5(b). 
Both $dp\pi$ and $dp\sigma$ levels shift downwards
but the shift in $dp\pi$ level is larger compared to $dp\sigma$
level because of the smaller O-2p contribution in $dp\pi$ level. As a result,
the $dp\pi$ and $dp\sigma$ levels lie roughly at the same energy. 
Thus the Mn d-d gap ($\Delta_{d-d}$) and O-2p Mn-d gap ($\Delta_{p-d}$) are of 
the same order (Fig. 5(b)).
Let us now see how this energy level diagram helps us to understand our DOS
results shown in Figs. 2-3. We first consider LSDA monoclinic case shown in 
Fig. 3a. We
see that Mn d-$t_{2g}$ states mainly contribute around E$_F$ while O-2p states
contribute well below E$_F$. This picture is similar to our energy level 
diagram shown in Fig. 5(a). When U is taken into account, we see from Fig. 3b 
that the $t_{2g}$ levels have been pushed down and O-2p contribution increases
near E$_F$. Thus the picture is similar to Fig. 5(b).
Fig. 2 can be understood in the similar fashion.

\subsection{X-ray Emission Spectra}
\subsubsection{Orthorhombic LiMnO$_2$}

To compare our work with the experimental data, we have calculated
Mn L$\alpha$ and O K$\alpha$ X-ray emission 
spectra (XES) for the antiferromagnetic orthorhombic structure 
using the LSDA and LSDA+U at U=8.0 eV which is shown in Fig. 6.
We have rescaled the intensity of calculated XES for better 
comparison with experimental result. In Mn $L_\alpha$ XES, we see that
the LSDA+U gives better agreement with experiment compared to the LSDA. 
In particular the region around peak 'B' of Mn spectrum produced by the LSDA+U
shows much better agreement with the experimental results \cite{gala}. 
These differences in the LSDA and LSDA+U spectra can be understood in terms of
DOS shown in Fig. 2, as the XES 
spectra have been calculated from Eqs. (1) and (2). In the LSDA spectrum, 
the feature 'A' corresponds to $t_{2g}$ DOS in energy range -7.0 eV
to -3.0 eV and feature 'B' corresponds to $e_{g}$ DOS in 
energy range -2.5 eV to -1.0 eV. In the LSDA+U spectrum, the 
features 'A' and 'B' correspond to $t_{2g}$ DOS in energy
range -7.0 to -3.0 eV and $e_{g}$ DOS in energy range -6.5 to -1.0 eV
respectively.
The peak 'B' is at -3.22 eV in the LSDA+U and at -2.56 eV in the LSDA while
the experimental peak is seen at 3.3 eV. 
In the LSDA spectrum of O $K_\alpha$, features 'D' and 'E' show the 
contribution of O(1)-2p and O(2)-2p states in energy ranges -7.0 to -3.0 eV
and feature 'F' shows the contribution of O(2)-2p states in energy range
-3.0 to 0.0 eV. 
However, in the LSDA+U, only features 'D' and 'E' are present. The feature 'D'
arises due to the contributions from O(1) and O(2) 2p-states in energy range 
-7.0 to -5.0 eV. The feature 'E' arises due to contributions from O(1) 2p-states
in energy range -5.0 to -1.5 eV and O(2) 2p-states in energy range -5.0 eV to
0.0 eV. The O(2)-2p DOS does not have sharp peaks in energy range -3.0 eV to 0.0 eV
and hence feature 'F' is absent in this case which is in agreement with
experimental work \cite{gala}.
The O K$\alpha$ XES peak
has maximum intensity at -4.2 eV in experimental work \cite{gala}, which
is seen in the LSDA calculation as well as in the LSDA+U calculation at -4.3 eV and at -4.1 eV respectively.

\subsubsection{Monoclinic LiMnO$_2$}

We have also calculated the Mn L$\alpha$ and O K$\alpha$ X-ray emission spectra 
for the AF3 monoclinic configuration by the LSDA and LSDA+U which is shown in
Fig. 7. We see that inclusion of U drastically changes the spectrum. This is
due to the change in the DOS caused by the inclusion of U. In Mn L$\alpha$
spectrum, the peak 'B' and feature 'C' have large intensity in the LSDA 
calculation as compared to the LSDA+U while the feature 'A' has lower 
intensity. This change in intensity shifts the highest peak structure from 
-1.6 eV ('B' peak in the LSDA) to -5.2 eV ('A' feature in the LSDA+U). 
In the LSDA spectrum, the feature 'A' corresponds to $t_{2g}$ as well
as $e_{g}$ DOS in energy range -7.0 to -3.0 eV and feature 'C' corresponds to 
$t_{2g}$ DOS in energy range -1.0 eV to 0.0 eV.  
The peak 'B' corresponds to $e_{g}$ DOS in energy
range -2.5 eV to -1.5 eV. 
In the LSDA+U spectrum, feature 'A' corresponds to $t_{2g}$ and $e_{g}$ DOS 
in energy range -6.5 eV to -5.0 eV and feature 'C' has mainly contribution of
$e_{g}$ DOS around E$_F$. The flat part of the spectrum between features 'A'
and 'C' has contribution of $t_{2g}$ and $e_{g}$ DOS in energy range -5.0 eV
to -1.0 eV.
In the O K$\alpha$ 
spectrum the peaks have lower intensity in the LSDA calculation compared to 
the LSDA+U calculation. The highest intensity is at -3.5 eV (feature 'D') in 
the LSDA calculation while it is at -1.7 eV in the LSDA+U. 
We note that the peaks 'E' and 'F' have been shifted to higher 
intensity in the LSDA+U calculation while feature 'D' remains absent due to 
change in DOS in energy range -7.0 eV to -5.0 eV as compared to the LSDA. 
The peaks 'E' and 'F' correspond to O-2p DOS in energy range -2.0 eV to -1.0 eV
and -1.0 eV to 0.0 eV respectively in the LSDA calculations. In the LSDA+U
spectrum, the peaks 'E' and 'F' have contribution of O-2p DOS in energy range
-5.0 eV to -1.0 eV and -1.0 eV to 0.0 eV respectively.
The inclusion of U changes the peak intensity for both Mn spectrum 
as well as oxygen spectrum. This effect of U in peak intensity is similar to 
the orthorhombic case but the change in peak intensity and the shift in highest
intensity peak is larger in this case. This difference arises because of the
difference in the DOS of both structures. In orthorhombic case, the $e_g$ and
$t_{2g}$ DOS are mainly distributed in energy range -5.0 to -1.5 eV and
-6.5 to -3.5 eV respectively. The $t_{2g}$ DOS contributes to the peak at $\approx$
-5.0 eV in XES spectrum and $e_{g}$ DOS contributes to the peak at -3.2 eV.
In case of monoclinic, the $e_g$ and $t_{2g}$ DOS are distributed in energy 
range -6.5 to -1.0 eV and a sharp peak near $E_F$. The peak at -5.2 eV has
contribution of $t_{2g}$ and $e_{g}$ DOS in energy range -6.5 eV to -5.0 eV
and the flat part of spectrum between energy
range -5.0 to -1.0 V has contribution of both $t_{2g}$ and $e_g$ DOS.
The peak near 0.0 eV is mainly contributed by $e_{g}$ DOS.
For monoclinic structure no X-ray emission experiment has been reported
till now and it
would be interesting to do such an experiment to verify our predictions.

\subsection{Effect of U on Mn Valency}

To see the effect of U on valency of Mn, we have examined the partial charges 
of Mn electrons in the LSDA and LSDA+U. Table II shows the partial
charges of Mn in the AFM orthorhombic structure for the LSDA and LSDA+U.
It is clear from Table II that the Mn 4$s$-states are fully ionized
in the LSDA as well as in the LSDA+U.  The partial charges
of majority and minority $p$-states are not affected by inclusion of U
while the partial charges in the majority Mn $d$-states have shifted from the
minority Mn $d$-states. Hence the total number of Mn-d electrons do not change 
much which shows that the inclusion of U does not affect Mn valency in this 
case. We have noticed the similar effect of inclusion of U on Mn valency in 
the monoclinic and rhombohedral structure.
The shift in charges occurs between the majority and the minority Mn-d electrons
after inclusion of U. Thus we do not expect the change in Mn valency due to
the inclusion of U. 

\section{Conclusions}

Our study using the LSDA+U method shows that the electronic structures of
orthorhombic, monoclinic and rhombohedral LiMnO$_{2}$ are significantly 
modified by inclusion of strong correlations. 
We have shown that the LiMnO$_{2}$ is an antiferromagnetic insulator by the LSDA+U 
for all structures while the LSDA predicts antiferromagnetic insulator for 
orthorhombic and monoclinic structures and FM metal for 
rhombohedral structure. 
The DOS calculations show a mixture of charge transfer and Mn 
d$\leftrightarrow$d excitation like 
insulator by the LSDA+U for monoclinic and orthorhombic LiMnO$_{2}$
in contrast to the LSDA result which shows a band insulator. On the
other hand, the LSDA+U calculation for the rhombohedral LiMnO$_{2}$ shows
it to be a Mott-insulator while the LSDA shows it to be a metal.
We also note a large enhancement of the O-2p states at the top of valence band
in the DOS of monoclinic and orthorhombic system by the LSDA+U.
The calculated X-ray emission photoelectron spectra for orthorhombic
and monoclinic structures show a decrease in peak intensity for Mn spectra 
and increase in peak intensity of oxygen spectra by the LSDA+U as compared
to the LSDA results. We find good agreement between the LSDA+U and experimental
results at U=8.0 eV with respect to Mn magnetic moment, nature of band gap and
X-ray emission photoelectron spectra for orthorhombic LiMnO$_{2}$.
The DOS calculations show that the strong correlations change the 
DOS features dramatically around the Fermi level for all structures. 
We find that while the inclusion of U increases Mn magnetic moments but
it does not affect Mn valency.

\begin{center}
\textbf{Acknowledgment}
\end{center}

It is a pleasure to thank Drs. Roy Benedek, M. K. Harbola, R. C. Budhani and 
S. Auluck for helpful
discussion. We are also thankful to V. R. Galakhov for providing his data of 
X-ray photoelectron spectrum on LiMnO$_{2}$. 
This work was supported by the Council of Scientific and Industrial
Research, New Delhi, under scheme No. 03(968)02/EMR-II.

\newpage

TABLE I. Band gap and magnetic moment ($m$) of Mn in the AFM orthorhombic LiMnO$_{2}$ and AF3 monoclinic and rhombohedral LiMnO$_{2}$ structure by the LSDA and LSDA+U at different U values (in eV).\\
\\

\begin{tabular}{|c l |l |l l l |l}\hline

	&
	&{\textsf{LSDA}}
	&
	&{\textsf{LDA+U}}
	&
\\\cline{4-6}
	&
	&
	&{\textsf{U=5.5 eV}}
	&{\textsf{U=6.9 eV}}
	&{\textsf{U=8.0 eV}}
\\\hline
	 {\textsf{Orthorhombic}}
	&{\textsf{  $m$ ($\mu_{B}$)}}
	&{\textsf{ 3.29}}
	&{\textsf{ 3.57}}
	&{\textsf{ 3.65}}
	&{\textsf{ 3.69}}
\\
	&{\textsf{Band Gap (eV)}}
	&{\textsf{ 0.30}}
	&{\textsf{ 1.50}}
	&{\textsf{ 1.78}}
	&{\textsf{ 1.95}}
\\\hline
	 {\textsf{Monoclinic}}
	&{\textsf{  $m$ ($\mu_{B}$)}}
	&{\textsf{ 3.29}}
	&{\textsf{ 3.67}}
	&{\textsf{ 3.74}}
	&{\textsf{ 3.79}}
\\
	&
	 {\textsf{Band Gap (eV)}}
	&{\textsf{ 0.61}}
	&{\textsf{ 1.49}}
	&{\textsf{ 1.60}}
	&{\textsf{ 1.80}}
\\\hline
	 {\textsf{Rhombohedral}}
	&{\textsf{  $m$ ($\mu_{B}$)}}
	&{\textsf{ 2.29}}
	&{\textsf{ 3.59}}
	&{\textsf{ 3.65}}
	&{\textsf{ 3.70}}
\\
	&{\textsf{Band Gap (eV)}}
	&{\textsf{ 0.00}}
	&{\textsf{ 0.30}}
	&{\textsf{ 0.65}}
	&{\textsf{ 0.85}}
\\\hline
\end{tabular}\\
\\
\\

TABLE II. Partial charges of Mn in the AFM orthorhombic LiMnO$_{2}$ by the LSDA and LSDA+U at different U values (in eV).\\
\\
\\

\begin{tabular}{|c l |l |l l l |l}\hline

	&
	&{\textsf{LSDA}}
	&
	&{\textsf{LDA+U}}
	&
\\\cline{4-6}
	&
	&
	&{\textsf{U=5.5 eV}}
	&{\textsf{U=6.9 eV}}
	&{\textsf{U=8.0 eV}}
\\\hline
	 {\textsf{Majority}}
	&{\textsf{  $s$}}
	&{\textsf{ 0.09}}
	&{\textsf{ 0.09}}
	&{\textsf{ 0.09}}
	&{\textsf{ 0.09}}
\\
	&{\textsf{$p$}}
	&{\textsf{ 3.07}}
	&{\textsf{ 3.08}}
	&{\textsf{ 3.08}}
	&{\textsf{ 3.08}}
\\
	&{\textsf{$d$}}
	&{\textsf{ 3.90}}
	&{\textsf{ 4.01}}
	&{\textsf{ 4.04}}
	&{\textsf{ 4.05}}
\\
	&{\textsf{Total}}
	&{\textsf{ 7.06}}
	&{\textsf{ 7.18}}
	&{\textsf{ 7.21}}
	&{\textsf{ 7.22}}
\\\hline
	 {\textsf{Minority}}
	&{\textsf{  $s$}}
	&{\textsf{ 0.07}}
	&{\textsf{ 0.08}}
	&{\textsf{ 0.08}}
	&{\textsf{ 0.08}}
\\
	&{\textsf{$p$}}
	&{\textsf{ 3.06}}
	&{\textsf{ 3.06}}
	&{\textsf{ 3.06}}
	&{\textsf{ 3.06}}
\\
	&{\textsf{$d$}}
	&{\textsf{ 0.64}}
	&{\textsf{ 0.47}}
	&{\textsf{ 0.42}}
	&{\textsf{ 0.39}}
\\
	&{\textsf{Total}}
	&{\textsf{ 3.77}}
	&{\textsf{ 3.61}}
	&{\textsf{ 3.56}}
	&{\textsf{ 3.53}}
\\\hline
\end{tabular}\\
\\
\\

TABLE III. Total energies ( with respect to rhombohedral LiMnO$_{2}$ in FM configuration ) of the orthorhombic, monoclinic and rhombohedral structures in ferromagnetic and antiferromagnetic configuration by the LSDA and LSDA+U at different U values (in eV).\\
\\

\begin{tabular}{|c l |l |l l l |l}\hline

	&
	&{\textsf{LSDA}}
	&
	&{\textsf{LDA+U}}
	&
\\\cline{4-6}
	&
	&
	&{\textsf{U=5.5 eV}}
	&{\textsf{U=6.9 eV}}
	&{\textsf{U=8.0 eV}}
\\\hline
	 {\textsf{Orthorhombic}}
	&{\textsf{AF (eV)}}
	&{\textsf{0.276}}
	&{\textsf{-1.007}}
	&{\textsf{-0.958}}
	&{\textsf{-1.299}}
\\
	&{\textsf{FM (eV)}}
	&{\textsf{0.577}}
	&{\textsf{-0.916}}
	&{\textsf{-0.886}}
	&{\textsf{-1.249}}
\\\hline
	 {\textsf{Monoclinic}}
	&{\textsf{AF (eV)}}
	&{\textsf{0.285}}
	&{\textsf{-0.696}}
	&{\textsf{-0.631}}
	&{\textsf{-0.812}}
\\
	&
	 {\textsf{FM (eV)}}
	&{\textsf{0.891}}
	&{\textsf{-0.496}}
	&{\textsf{-0.451}}
	&{\textsf{-0.640}}
\\\hline
	 {\textsf{Rhombohedral}}
	&{\textsf{AF (eV)}}
	&{\textsf{0.521}}
	&{\textsf{-0.592}}
	&{\textsf{-0.588}}
	&{\textsf{-0.581}}
\\
	&{\textsf{FM (eV)}}
	&{\textsf{ 0.00}}
	&{\textsf{ 0.00}}
	&{\textsf{ 0.00}}
	&{\textsf{ 0.00}}
\\\hline
\end{tabular}

\newpage
\begin{figure}
$\bf{FIGURE}$ $\bf{CAPTIONS}$

\caption{\label{fig1} Antiferromagnetic ordering for orthorhombic LiMnO$_{2}$ in two planes z=0.4 and z=0.6. Only Mn atoms are shown.}
\caption{\label{fig2} Total DOS and the partial DOS of majority spin and minority spin Mn 
d-t$_{2g}$ bands and Mn d-e$_{g}$ bands for AFM orthorhombic LiMnO$_{2}$
(a) for the LSDA and (b) for the LSDA+U. The LSDA+U DOS corresponds to U=8.0 eV.  The bottom two panels show the partial DOS of majority spin O-2p bands at two 
different planes. O1 and O2 correspond to 2b site at z=0.13 and z=0.60 
respectively in P$_{mmn}$ space group. 
The dashed vertical line denotes the Fermi level E$_{F}$.
The DOS for the majority spin
is shown on the upside and DOS for the minority spin, on the downside.}
\caption{\label{fig3} Total DOS and the partial DOS of majority spin and minority spin Mn 
d-t$_{2g}$ bands and Mn d-e$_{g}$ bands for AF3 monoclinic LiMnO$_{2}$
(a) for the LSDA and (b) for the LSDA+U (U=8.0 eV). 
The bottom panel shows the partial DOS of majority spin O-2p bands. 
The dashed vertical line denotes the Fermi level E$_{F}$.
The DOS for the majority spin
is shown on the upside and DOS for the minority spin, on the downside.}
\caption{\label{fig4} Total DOS for AF3 and FM rhombohedral LiMnO$_{2}$ (a) for the LSDA and (b) for the LSDA+U (U=8.0 eV).} 
\caption{\label{fig5} Schematic energy level diagram of molecular orbitals for MnO$_{6}$ octahedron in LiMnO$_{2}$. (a) and (b) show the rough estimate for the LSDA and LSDA+U respectively.}
\caption{\label{fig6} X-ray emission photoelectron spectra of Mn L$\alpha$ and O K$\alpha$ using the LSDA and LSDA+U and experimental result (Ref. 12) for orthorhombic case. The solid and dashed curves show the LSDA and LSDA+U (U=8.0 eV) results respectively and the dots, experimental result.} 
\caption{\label{fig7} X-ray emission photoelectron spectra of Mn L$\alpha$ and O K$\alpha$ using the LSDA and LSDA+U for monoclinic case. The solid line and dash line curves are showing the LSDA and LSDA+U (U=8.0 eV) results respectively. Upper panel shows Mn L$\alpha$ spectrum and lower panel shows O K$\alpha$ spectrum.} 
\end{figure}

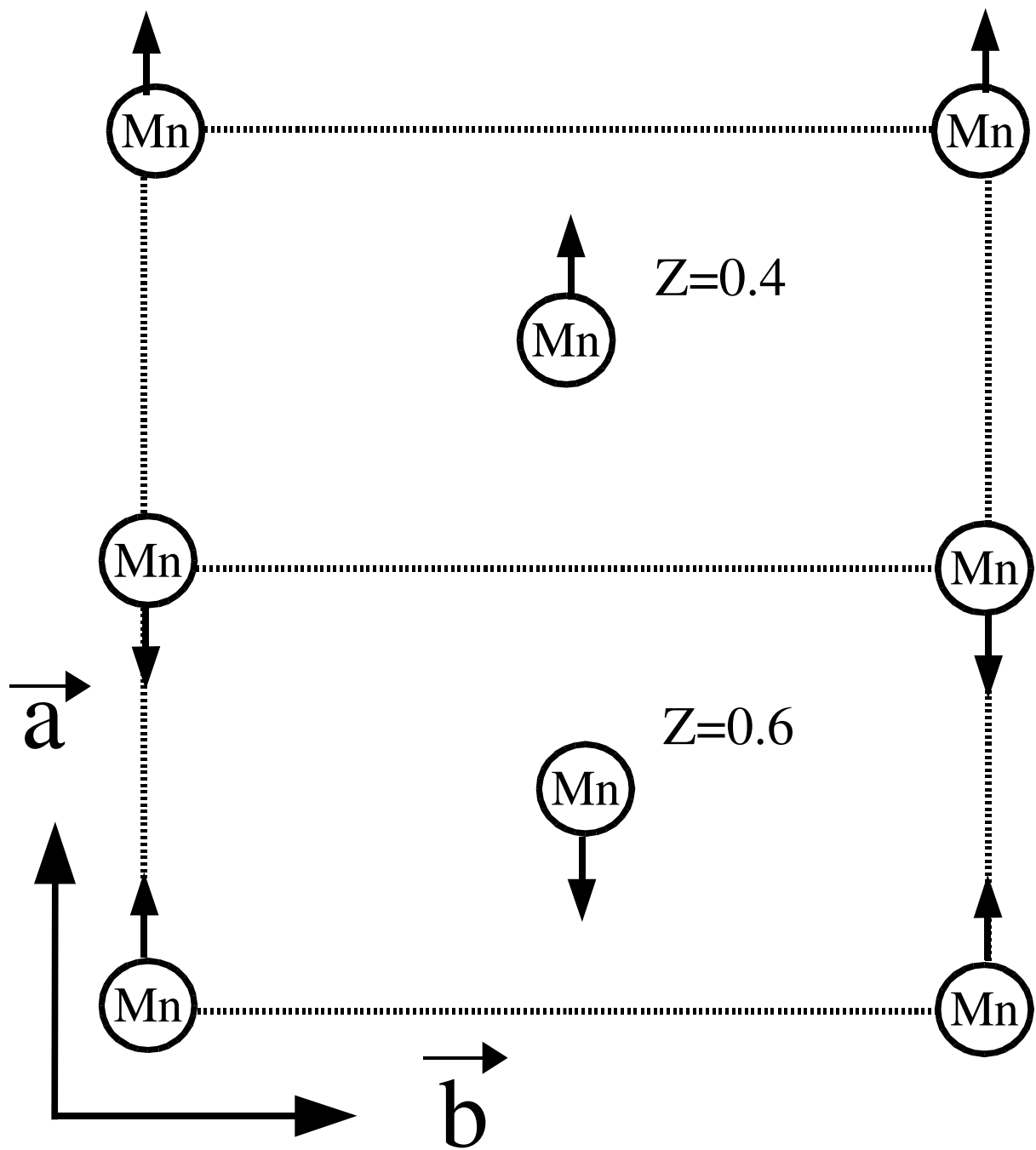
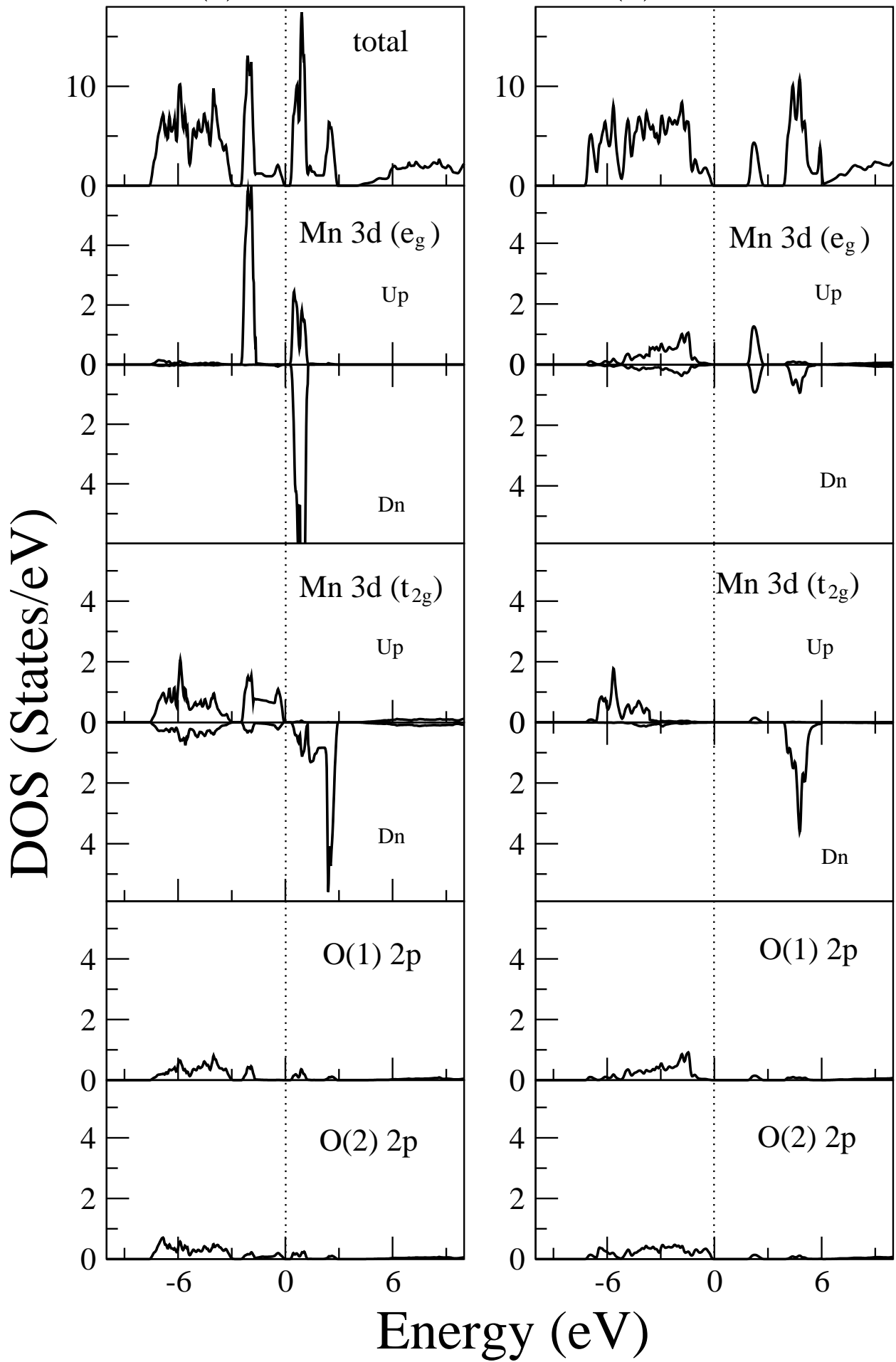
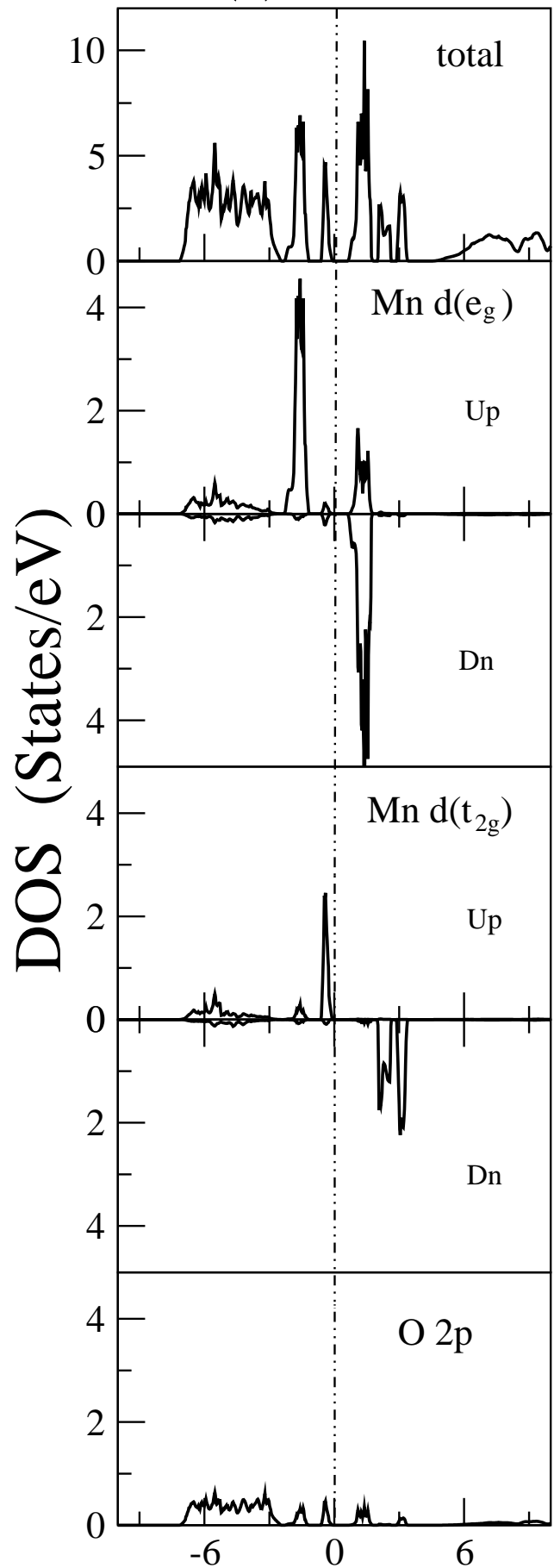
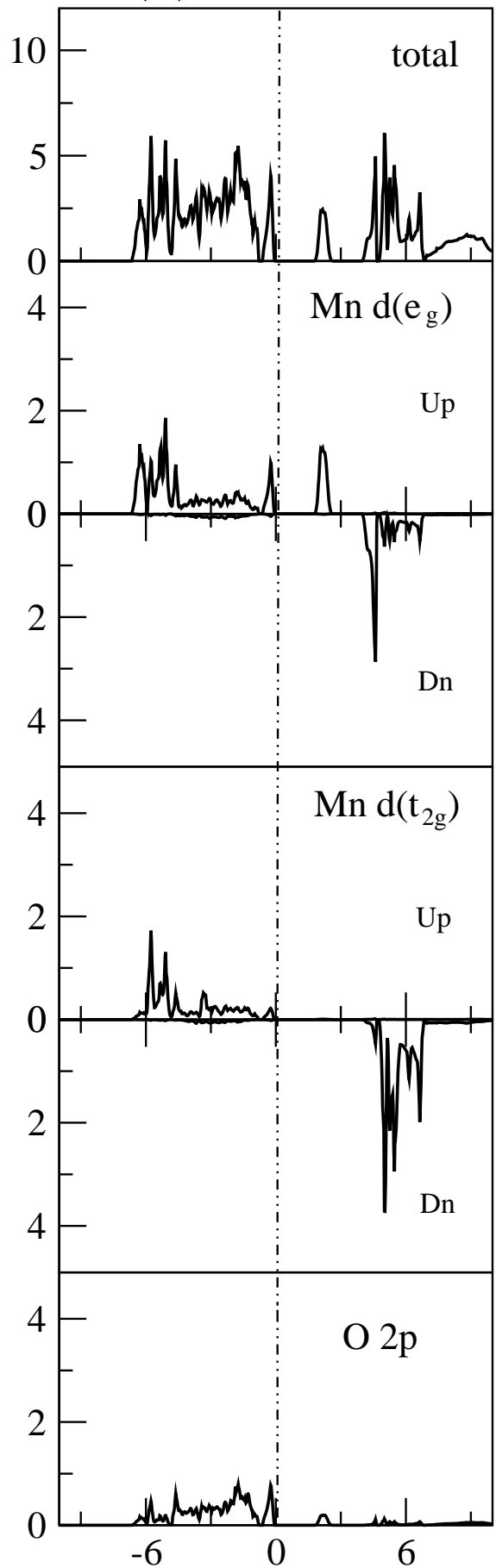
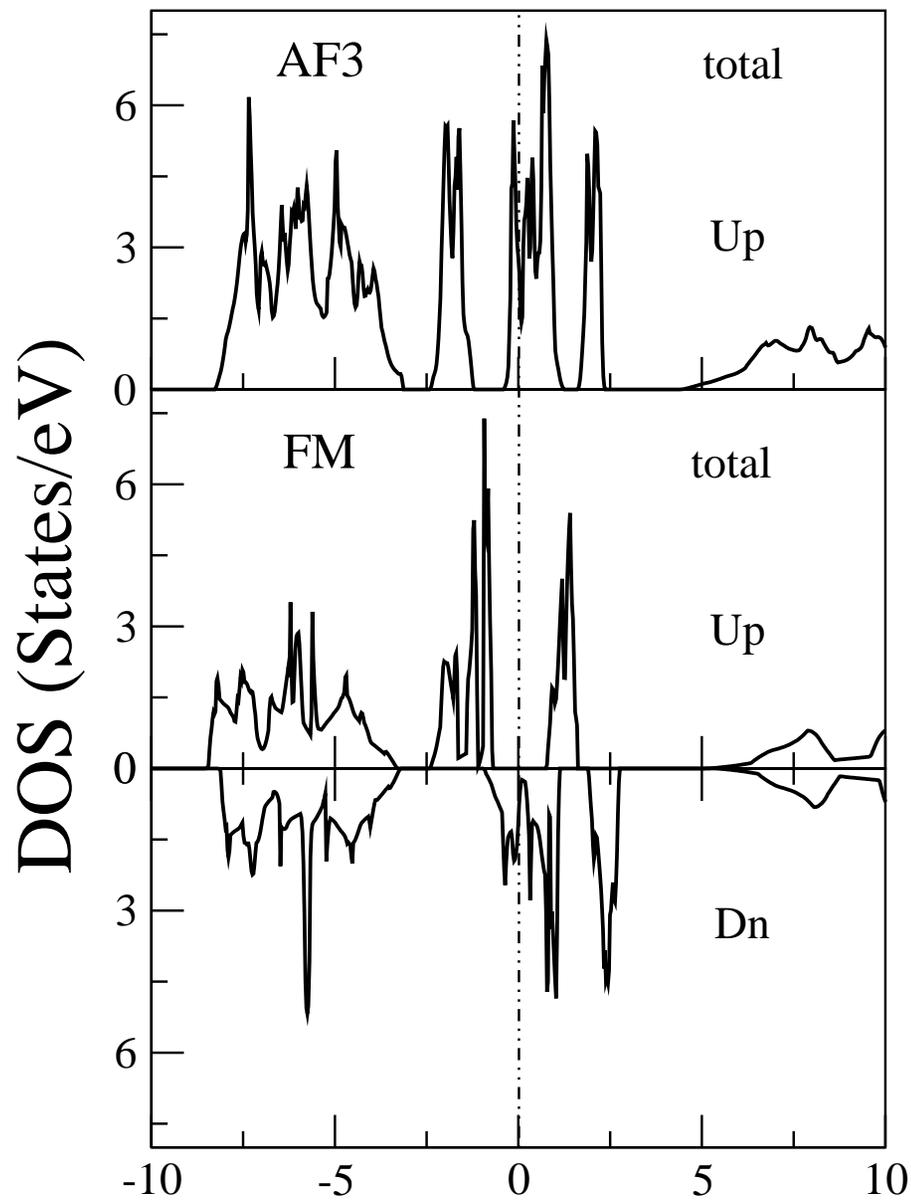
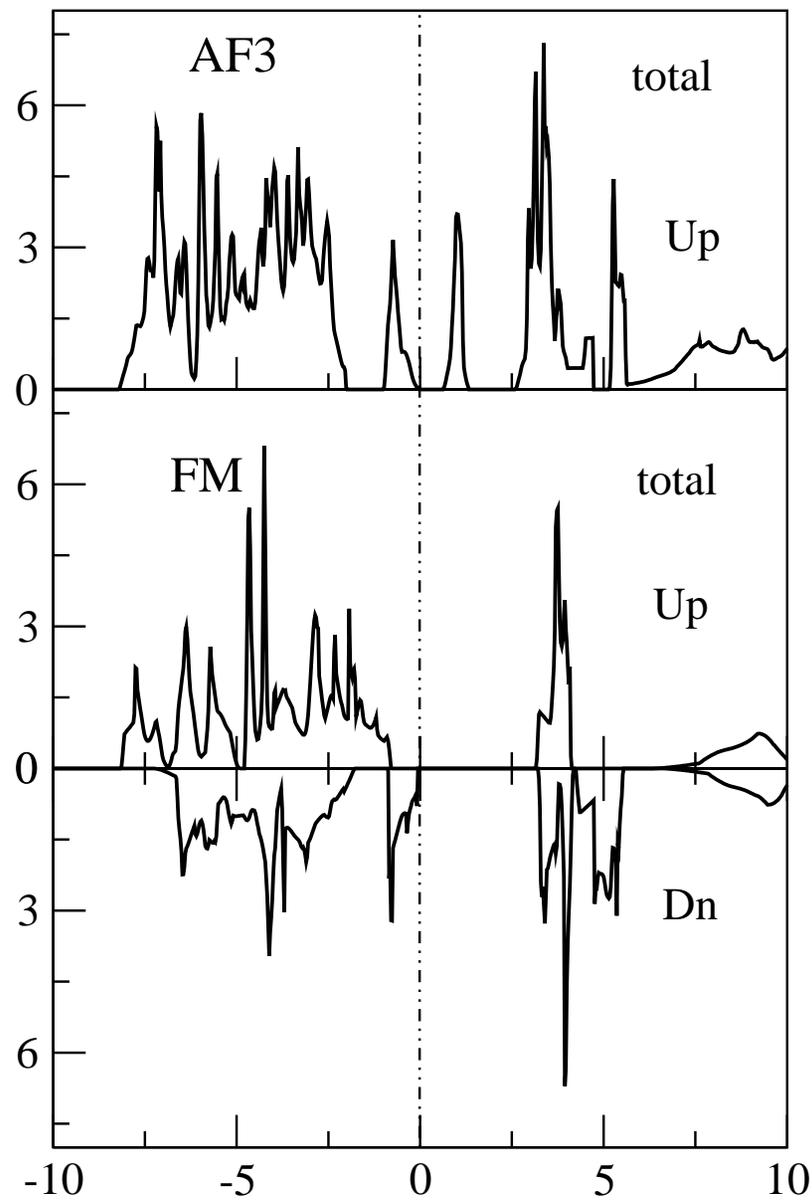
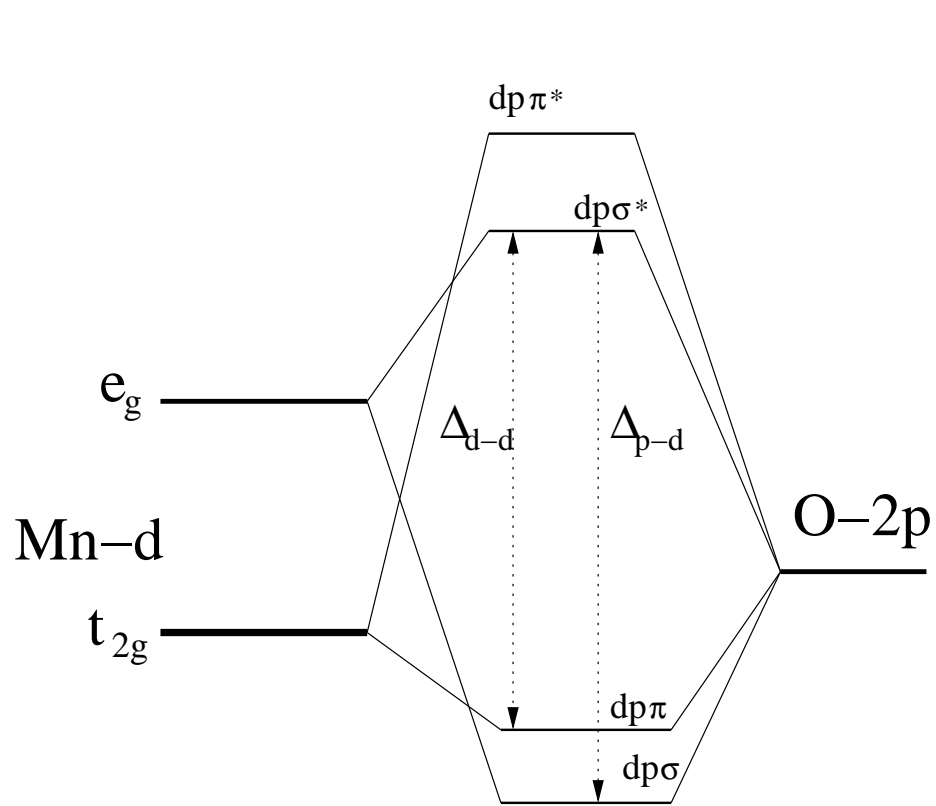
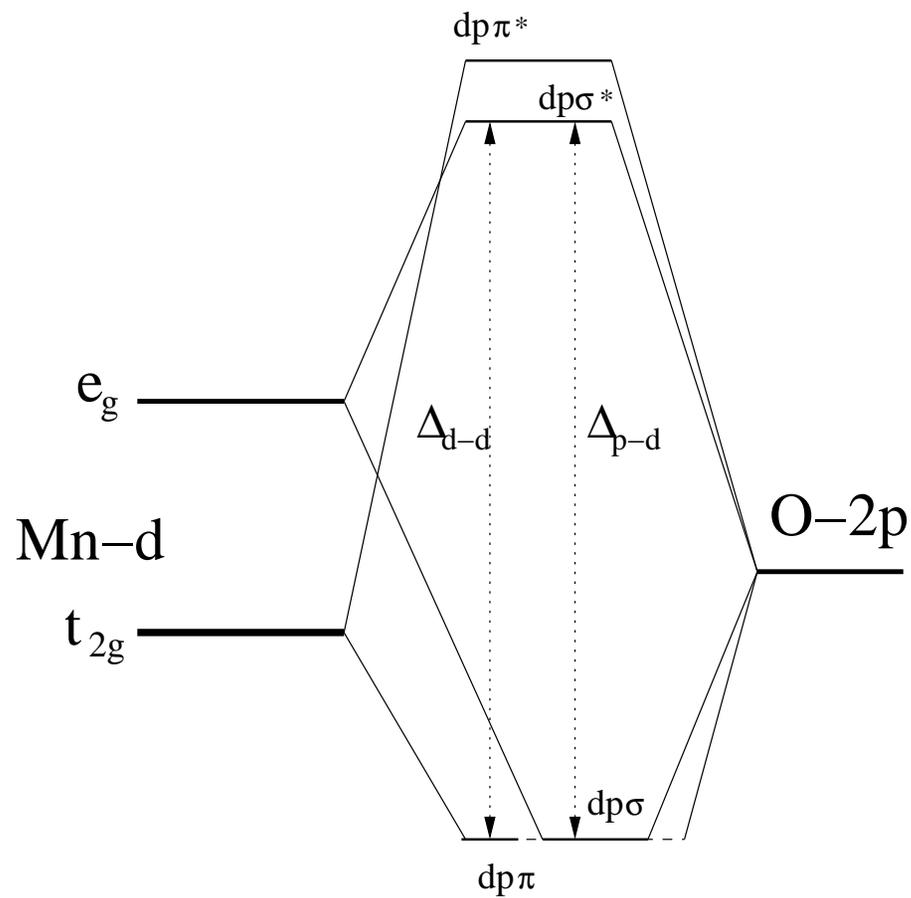
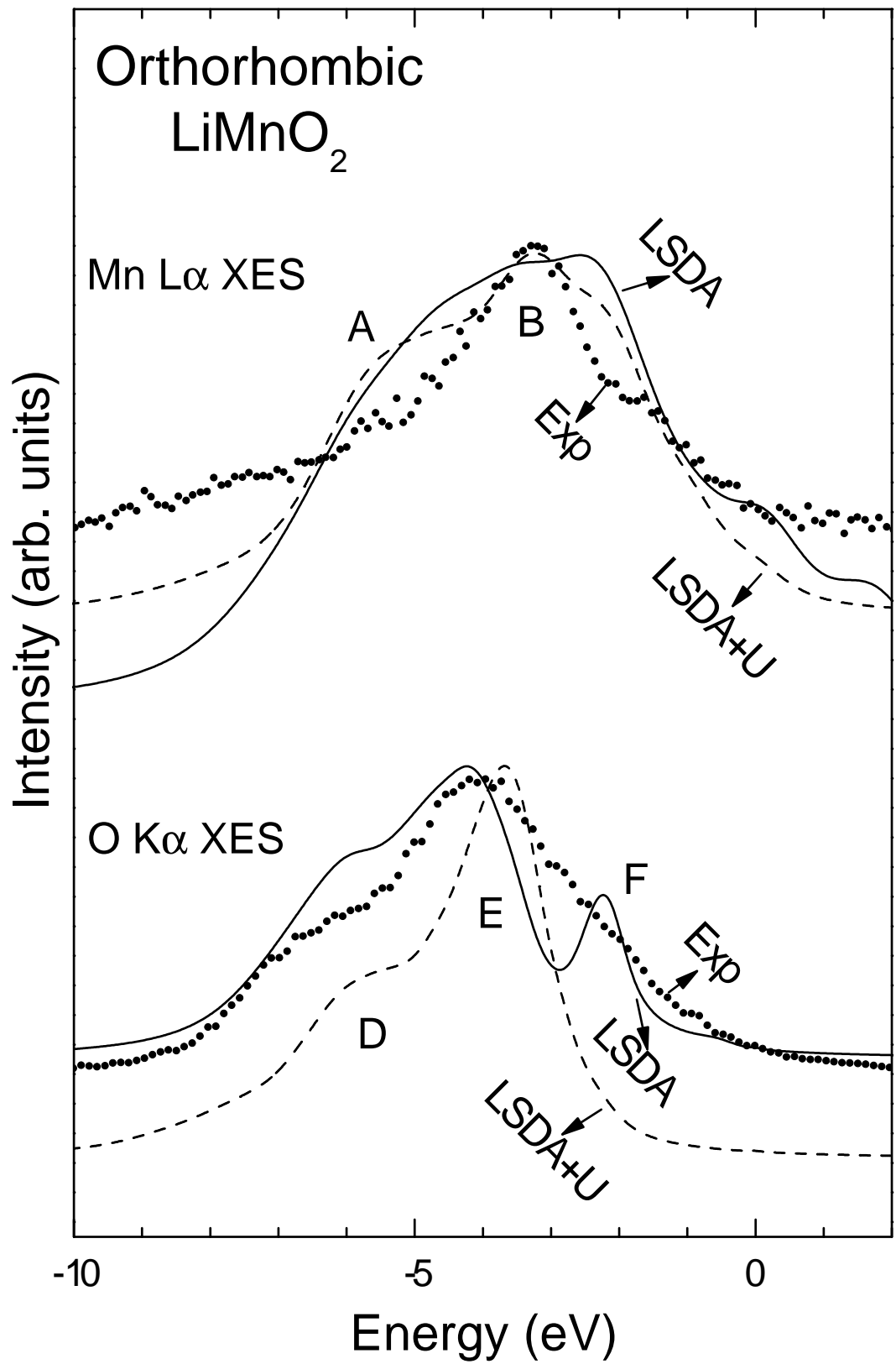
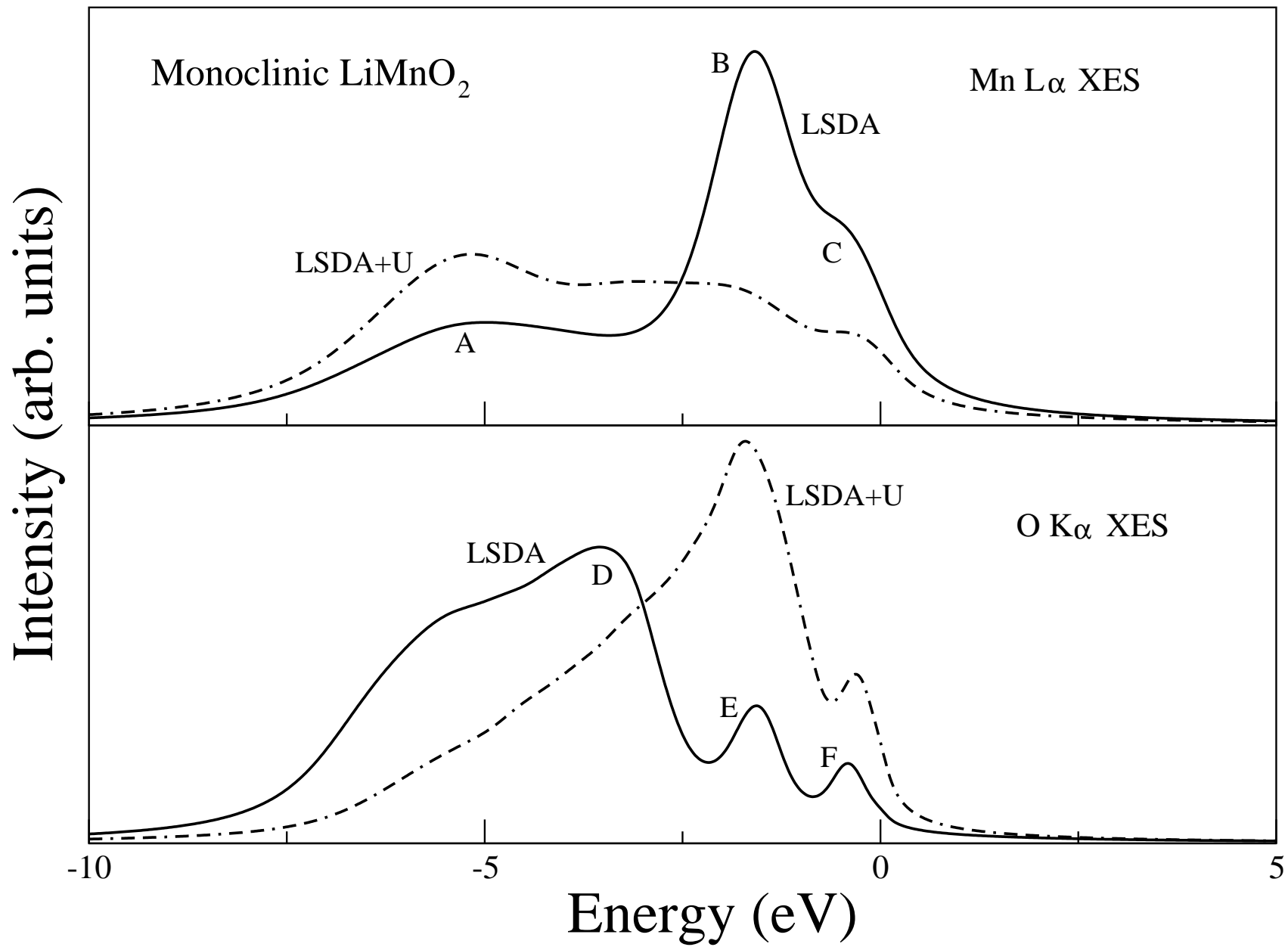
\end{document}